\documentclass[10pt,letterpaper]{article}

\usepackage{opex3}
\usepackage{amsmath}

\usepackage{graphicx}
\graphicspath{{pict/}{}}
\usepackage{bm}

\newcommand{\be}{\begin{equation}}
\newcommand{\ee}{\end{equation}}

\newcommand\pictc[5]{\begin{figure}
                       \centerline{
\includegraphics[width=#1\columnwidth,height=0.7\textheight,keepaspectratio]{#3}}
\protect\caption{\protect\label{fig:#4} #5}
                    \end{figure}            }
\newcommand\pict[4][0.78]{\pictc{#1}{!tb}{#2}{#3}{#4}}
\newcommand\rpict[1]{\ref{fig:#1}}

\newcommand\leqt[1]{\protect\label{eq:#1}}
\newcommand\reqtn[1]{\ref{eq:#1}}
\newcommand\reqt[1]{(\reqtn{#1})}

\newcounter{Fig}

\begin{document}

\title{Slow light with flat or offset band edges in multi-mode fiber with two gratings}

\author{Andrey A. Sukhorukov$^1$, C. J. Handmer$^2$,
        C. Martijn de Sterke$^{2}$  and M.~J.~Steel$^{3,2}$}

\address{
ARC Center of Excellence for Ultrahigh-bandwidth Devices for Optical Systems (CUDOS),\\
$^1$~Nonlinear Physics Centre,
Research School of Physical Sciences and Engineering,
Australian National University, Canberra, ACT 0200, Australia\\
$^2$ School of Physics, University of Sydney, Camperdown, NSW 2006, Australia\\
$^3$ RSoft Design Group, Inc., 65 O'Connor St, Chippendale, NSW 2008, Australia}

\email{ans124@rsphysse.anu.edu.au}

\begin{abstract}
We consider mode coupling in multimode optical fibers using either two Bragg
gratings or a Bragg grating and a long-period grating. We show that the
magnitude of the band edge curvature can be controlled leading to a flat,
quartic band-edge or to two band edges at distinct, nonequivalent $k$-values,
allowing precise control of slow light propagation.
\end{abstract}


\noindent The ability to control the propagation of waves using the strong
geometrical dispersion in photonic structures opens new opportunities for
manipulating optical pulses. In particular, the regime of slow light can be
achieved when the wavelength is tuned close to the edges of a photonic
band-gap, as was demonstrated both in one-dimensional \cite{Mok:2006-775:NAPH} and
two-dimensional \cite{Letartre:2001-2312:APL} geometries. The properties of slow-light modes
strongly depend on the shape of the dispersion curve: if the curvature at the
band edge is eliminated, to lowest order $\omega-\omega_E\propto k^4$, where
$\omega_E$ is the band-edge frequency and $k$ the wavenumber. This results in a flat band-edge for
which the group velocity $v_g\propto k^3\propto (\omega-\omega_E)^{3/4}$. In
addition, the curvature can be made to change sign so that the band edges
appear at distinct non-equivalent $k$-values.

A flat band edge can lead to high energy densities and to an increased density
of states at frequencies near the band edge~\cite{Figotin:2005-36619:PRE,
Ibanescu:2004-63903:PRL}, with associated changes in the radiative properties of
sources. The increase in the energy density is associated with fact that for a
given, low group velocity pulse, the range of $k$ values in a quartic band is
larger than in a quadratic band, so that the pulse can be more strongly
compressed in space. The increased density of states directly follows from the
flattened shape of the dispersion relation.  Slow modes with non-vanishing
phase velocities can also be used to create high-$Q$
resonators~\cite{Ibanescu:2005-552:OL}. It was earlier shown that such
dispersion characteristics can be achieved in extended multi-layer anisotropic
structures~\cite{Figotin:2005-36619:PRE, Cao:2006-ME16:ProcSL} or omni-guide
fibers~\cite{Ibanescu:2004-63903:PRL, Ibanescu:2005-552:OL}. Flexible control
of higher-order dispersion is also possible in photonic-crystal
waveguides~\cite{Petrov:2004-4866:APL, Mori:2005-9398:OE}.

\pict{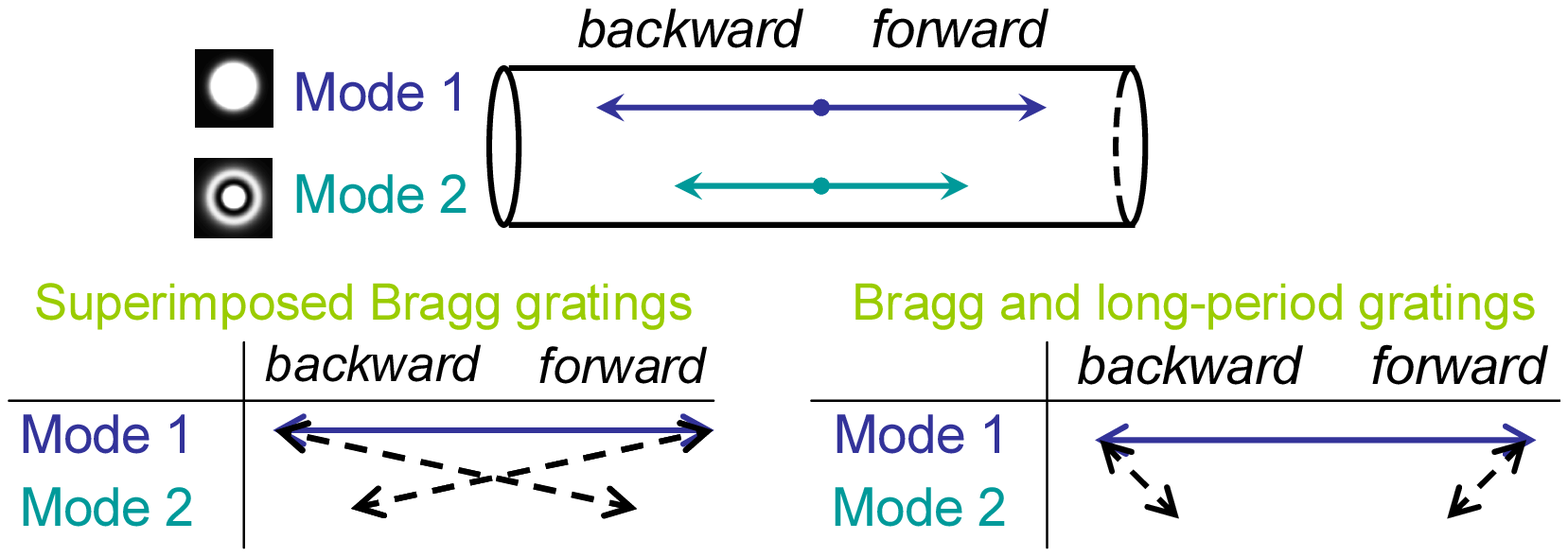}{sketch}{
Schematic of the coupling between the forward and backward propagating modes in a guided-wave structure using two superimposed Bragg gratings or Bragg and long-period gratings.
Top: schematic of mode wavenumbers.
Bottom: Solid and dashed arrows indicate mode coupling by individual gratings.
}

In this work, we suggest a generic approach for the engineering of flat and
offset band-edges in conventional fibers based on the grating coupling of
forward- and backward-propagating modes of two different symmetries, {\sl
i.e.}, 4 modes in total. In the most common geometry in such structures, a
Bragg grating is used to couple the fundamental mode to itself, leading to
Bragg reflection and the opening of a (one-dimensional) photonic bandgap. The
manipulation of this gap requires an additional degree of freedom, which is
obtained by coupling two modes (such as the fundamental mode and a higher-order mode) using an
additional grating, either a Bragg grating, or a long-period grating (see
Fig.~\rpict{sketch}). This is somewhat similar to earlier proposals
\cite{Figotin:2005-36619:PRE, Cao:2006-ME16:ProcSL}, however in that work, a
thin-film geometry in which the additional field has a different polarization
was considered. Our proposal is based on effects which cannot be achieved with
a dual Bragg grating~\cite{Kang:2002-1049:ELL} optimized only for a single
guided mode. The suggested design does not require the use of anisotropic
layers~\cite{Figotin:2005-36619:PRE, Cao:2006-ME16:ProcSL} or a special fiber
configuration~\cite{Ibanescu:2004-63903:PRL, Ibanescu:2005-552:OL}, being
readily accessible with established grating-writing methods in conventional
optical fibers.

We consider an optical fiber supporting two different modes, ${\cal M}_1$ and
${\cal M}_2$ (for example, LP$_{01}$ and LP$_{02}$). In the vicinity of a
chosen frequency $\tilde{\Omega}$, their dispersion relations are given by $\Omega_1 =
\pm V_1 (K_1 - \tilde{K}_{1})$, $\Omega_2 = \pm V_2 (K_2 - \tilde{K}_{2})$.
Here the $\Omega_j$ are the frequency detunings from $\tilde{\Omega}$, $K_j$
are corresponding wavenumber shifts, $V_j$ are the (positive) group velocities,
and $\tilde{K}_j$ ($>0$) are reference wavenumbers at $\tilde{\Omega}$. The signs $+$
and $-$ correspond to forward and backward propagating waves, respectively.

As mentioned, we introduce a grating supporting the simultaneous resonant
coupling of two pairs of forward- and backward-propagating modes ${\cal M}_1$
and ${\cal M}_2$.
This can be achieved with a superimposed modulation of the refractive index
along the fiber
\begin{equation} \leqt{grating}
   \Delta n(x,y,z) = \Delta n_1 \cos\left[\kappa_1 z +\phi_1 \right] R(x,y)+
   \Delta n_2 \cos\left[ \kappa_2 z+\phi_2\right] R(x,y),
\end{equation}
where the $R(x,y)$ defines the photosensitive
cross-section of the fiber. We choose $\kappa_1 = 2 \tilde{K}_1 + \delta_1$ and
$\kappa_2 = \tilde{K}_1 + \sigma \tilde{K}_2 + \delta_2$. Here $\delta_j$ are
small detunings from the two resonances, and $\sigma = \pm 1$. In both cases a
Bragg grating with wavevector $\kappa_1$ couples the forward- and
backward-propagating ${\cal M}_1$ mode. When $\sigma = +1$ an additional Bragg
grating couples the counter-propagating modes ${\cal M}_1$ and ${\cal M}_2$,
whereas for $\sigma = -1$ the co-propagating modes are coupled by a long-period
grating. The modulation amplitudes $\Delta n_1$ and $\Delta n_2$ define the
coupling strengths between the ${\cal M}_1$-${\cal M}_1$ and ${\cal
M}_1$-${\cal M}_2$ modes, respectively. Whereas it was previously found that
multi-mode coupling and conversion with superimposed Bragg gratings can give
rise to special features in the reflection and transmission spectra~\cite{
Erdogan:1996-296:JOSA, Case:1975-724:JOS, Mizrahi:1993-1513:JLT,
Yariv:1998-1835:OL, Lee:2001-1176:JOSA}, we demonstrate below that the
suggested superlattice modulation configuration enables new possibilities for
dispersion control of band-edge slow-light states.

We assume that the structure is much longer than the beat length between the
two modes, $L \gg |\tilde{K}_2 \pm \tilde{K}_1|^{-1},\, |\tilde{K}_1|^{-1},\,
|\tilde{K}_2|^{-1}$, which is satisfied under typical conditions. Then, the pulse
propagation can be modeled by the coupled-mode equations~\cite{
Erdogan:1996-296:JOSA, Mizrahi:1993-1513:JLT} for the envelopes of the forward
and backward propagating modes.
Writing the electric field as
\begin{align}
   \mathbf{E}(x, y, z, t)
   = \{ & \mathbf{g}_1(x,y) [ u_1(z,t) \exp(i \tilde{K}_1 z) + w_1(z,t) \exp(-i \tilde{K}_1 z) ]
    \nonumber \\
   + & \mathbf{g}_2(x,y) [ u_2(z,t) \exp(i \sigma \tilde{K}_2 z) + w_2(z,t)
       \exp(-i \sigma \tilde{K}_2 z) ] \}
       \exp(-i \tilde{\Omega} t) + \text{c.c.},
\end{align}
where $\mathbf{g}_{1,2}(x,y)$ are transverse profiles for the two modes and c.c. denotes the complex
conjugate, we see that the forward-propagating envelopes are $u_1$ and $u_2$ for $\sigma=+1$ or $u_1$ and $w_2$ for $\sigma=-1$,
and the backward-propagating envelopes are $w_1$ and $w_2$ for $\sigma=+1$ or $w_1$ and $u_2$ for $\sigma=-1$.
We choose the scaling where the normalized intensities define the energy density of modes per unit length of the fiber, and obtain the
dimensionless equations
\begin{equation} \leqt{CM}
   \begin{array}{l} {\displaystyle
      i \frac{\partial u_1}{\partial t}
      + i V_1 \frac{\partial u_1}{\partial z}
      + \rho_1 w_1 \exp( i \delta_1 z + i \phi_1)
      + \rho_2 w_2 \exp( i \delta_2 z + i \phi_2 )
      = 0,
   } \\*[9pt] {\displaystyle
      i \frac{\partial w_1}{\partial t}
      - i V_1 \frac{\partial w_1}{\partial z}
      + \rho_1 u_1 \exp( - i \delta_1 z -i \phi_1)
      + \rho_2 u_2 \exp( - i \delta_2 z -i \phi_2)
      = 0,
   } \\*[9pt] {\displaystyle
      i \frac{\partial u_2}{\partial t}
      + i \sigma V_2 \frac{\partial u_2}{\partial z}
      + \rho_2 w_1 \exp( i \delta_2 z +i \phi_2)
      = 0,
   } \\*[9pt] {\displaystyle
      i \frac{\partial w_2}{\partial t}
      - i \sigma V_2 \frac{\partial w_2}{\partial z}
      + \rho_2 u_1 \exp( - i \delta_2 z -i \phi_2)
      = 0.
   } \end{array}
\end{equation}
Here the (real-valued) grating strengths $\rho_j$ are approximately given by
\begin{equation}
   \rho_j = \frac{\pi \Delta n_j}{\lambda_0} \int \int_{\infty}^{\infty} \mathrm{d}x \, \mathrm{d}y  \,
             R(x,y) \mathbf{g}_1(x,y) \cdot \mathbf{g}_j(x,y),
\end{equation}
with $\lambda_0=2\pi c/\tilde{\Omega}$.
Note that the {\em values of $\rho_j$ can be chosen independently} by selecting the strength of corresponding grating
modulations and that Eqs.~\reqt{CM} satisfy the energy-conservation condition, $\partial( |u_1|^2 + |w_1|^2 + |u_2|^2 + |w_2|^2) / \partial t =
\partial( - V_1 |u_1|^2 + V_1 |w_1|^2 - \sigma V_2 |u_2|^2 + \sigma V_2 |w_2|^2) / \partial z$.

The dispersion relation of the system can be calculated by analyzing the eigenmodes of Eqs.~\reqt{CM} in the form,
\begin{align} \leqt{Bloch}
      u_1 &= U_1 \exp\left[ i (k + \delta_1/2) z - i \omega t + i \phi_1 / 2 \right], \nonumber \\
      w_1 &= W_1 \exp\left[ i (k - \delta_1/2) z - i \omega t - i \phi_1 / 2 \right], \nonumber \\
      u_2 &= U_2 \exp\left[ i (k - \delta_1/2 + \delta_2) z - i \omega t - i \phi_1 / 2 + i \phi_2  \right], \nonumber \\
      w_2 &= W_2 \exp\left[ i (k + \delta_1/2 - \delta_2) z - i \omega t  + i \phi_1  / 2 - i \phi_2  \right].
\end{align}
We present the eigenmodes as sums of symmetric and antisymmetric vectors, $U_j = F_{j+} + F_{j-}$ and $W_j = F_{j+} - F_{j-}$ for $j = 1,2$. After
substituting these expressions into Eqs.~\reqt{CM}, and obtaining the eigenmode equations, we find that they can be formulated in the matrix form
\begin{equation} \leqt{BlochEqSymm}
   M_+
   \left( \begin{array}{cccc}
          F_{1+} \\ F_{2+}
   \end{array} \right)
   =
   k
   \left( \begin{array}{cccc}
          F_{1-} \\ F_{2-}
   \end{array} \right)
   , \quad
   M_-
   \left( \begin{array}{cccc}
          F_{1-} \\ F_{2-}
   \end{array} \right)
   =
   k
   \left( \begin{array}{cccc}
          F_{1+} \\ F_{2+}
   \end{array} \right) ,
\end{equation}
where
\begin{equation} \leqt{BlochMatrS}
   M_\pm =
   \left( \begin{array}{cc}
         (\omega \pm \rho_1) / V_1 - \delta_1 / 2 , & \pm \rho_2 / V_1 \\
         \pm \rho_2 \sigma / V_2, & \omega \sigma / V_2 + \delta_1 / 2 - \delta_2
   \end{array} \right) . \quad
\end{equation}
Therefore, $k^2$ are eigenvalues of a square $2 \times 2$ real-valued matrix $M
= M_{+} M_{-}$. Thus there appear four branches of dispersion curves
$\omega_j(k)$, with symmetry $\omega_j(k) = \omega_j(-k)$. For convenience, we
label the branches such that $\omega_1 \le \omega_2 \le \omega_3 \le
\omega_4$.  From the form of the matrices $M_{\pm}$, we observe
that the dispersion does not depend on the grating phases $\phi_j$, and
identify the symmetry property $\omega_j(k; \delta_1, \delta_2, \rho_1,
\rho_2, V_1, V_2) = \rho_1 \omega_j(k V_1/ \rho_1; 0, \delta_2 V_1/ \rho_1 - \delta_1 V_1 [1
+ \sigma V_1 / V_2] / [2 \rho_1], 1, \rho_2 / \rho_1, 1, V_2/V_1 ) + \delta_1 V_1 / 2$.
Then, in the analysis and figures below we take $\delta_1 = 0$, $V_1=1$ and $\rho_1 = 1$, since for
other values the dispersion curves can be obtained by simple scaling.

Away from the grating resonances, $\omega_j(k\rightarrow \pm \infty) \simeq \pm
V_{1,2} k$, {\sl i.e.}, there are two pairs of branches with positive and
negative frequency detunings. Since all $\omega_j(k)$ are real for real $k$,
branches cannot merge or disappear. However, for some frequencies all $k$
can become imaginary, such that the propagation of linear waves is suppressed
due to Bragg reflection.
Such a photonic band-gap appears between the second and third branches and
due to the symmetry of the dispersion curves, the band-gap edges appear at
pairs of dispersion points labeled $(\pm k_2, \omega_2^{(b)})$ (lower) and $(\pm k_3,
\omega_3^{(b)})$ (upper edge). The wavenumbers $\pm k_2$ and $\pm k_3$ define
the corresponding phase velocity of slow-light modes.

\pict{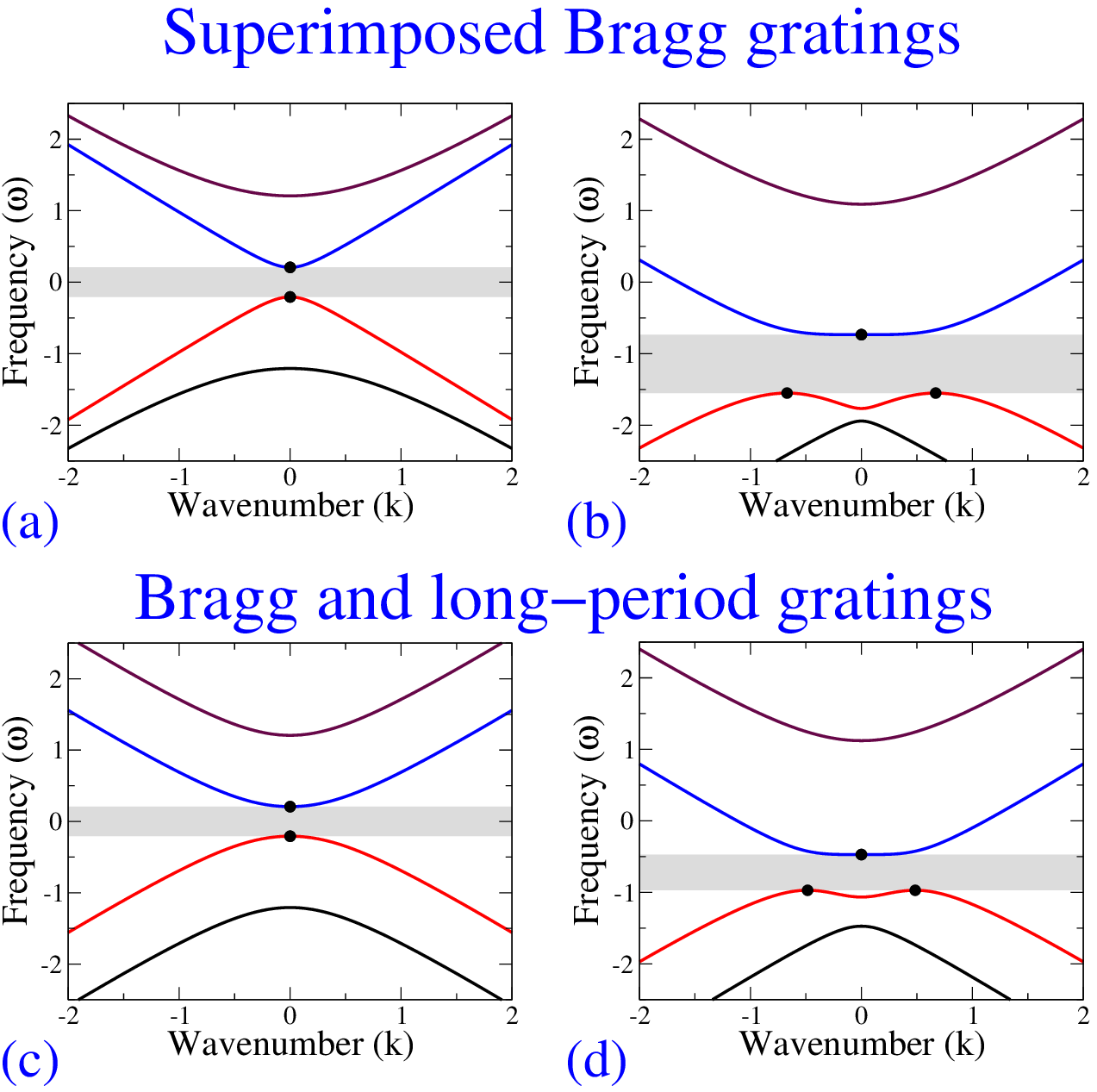}{dispersion}{Dispersion relation for structures based on
(a), (b)~two Bragg gratings ($\sigma=+1$) and (c), (d)~Bragg and long-period
gratings ($\sigma=-1$). Shading marks the band-gap, and filled circles indicate
the band edges. For all the plots, $V_1 = 1$, $V_2 = 0.95$, $\rho_1 = 1$,
$\rho_2 = 0.5$, $\delta_1 = 0$. Grating detunings are (a)~$\delta_2 = 0$,
(b)~$\delta_2 = -1.765$, (c)~$\delta_2 = 0$, (d)~$\delta_2 = 0.995$. }

\pict{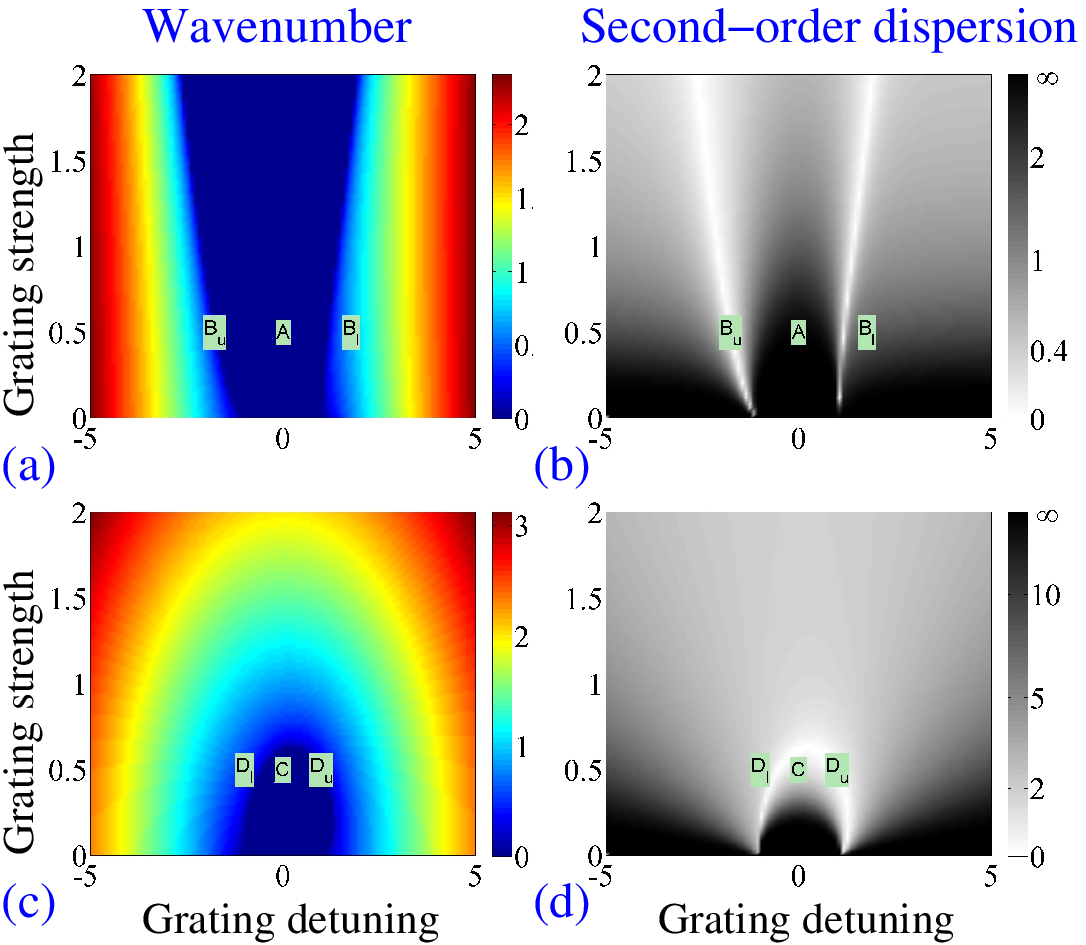}{tuning}{
Tuning characteristics of slow-light modes for (a),(b)~coupled Bragg gratings ($\sigma=1$) and (c),(d)~coupled Bragg and long-period gratings ($\sigma=-1$).
Shown are the dependencies of the absolute values of the band-edge (a,c)~wavenumber and (b,d)~group-velocity dispersion
on the grating strength ($\rho_2$) and detuning ($+\delta_2$ at upper or $-\delta_2$ at lower gap edges).
Marked points A--D correspond to the dispersion plots in Figs.~\rpict{dispersion}(a)--(d), respectively, where the subscripts $u$ and $l$ indicate the upper and lower gap edges (points A and C do not have subscripts as they coincide for both band-edges).
Note that in (a) and (c), the band-edge wavenumbers are exactly zero throughout the dark blue region.
Normalized parameters are the same as in Fig.~\rpict{dispersion}.
}

To determine the features of band-edge dispersion, we first analyze the point
with $k = 0$. Then, Eqs.~\reqt{BlochEqSymm} decouple, and admit solutions in
the form of purely symmetric or antisymmetric modes when ${\rm Det}\{M_{\pm}[\omega_j(k=0)] \} = 0$, the solutions of which
satisfy
\begin{equation} \leqt{edgesol}
   (\omega_j \pm 1)(\omega_j / V_2 - \sigma \delta_2) = \rho_2^2 / V_2.
\end{equation}
The dispersion points $\omega_{j}(k=0)$ at $j=2$ or $j=3$ correspond to the
edges of the photonic band-gap if the second eigenvalue of $M[\omega_j(k=0)]$
is negative, i.e. if no other propagating modes exist in the spectral region
around $\omega_{j}(k=0)$.
This condition is satisfied if
\begin{equation}
    \eta_j = (\omega_j / V_2 - \sigma \delta_2)^2 + (\omega_j^2 - 1) - 2 \sigma \rho_2^2 / V_2 < 0 .
\end{equation}
Therefore, $\omega_j^{(b)} = \omega_j(k=0)$ and $k_j = 0$ if $\eta_j < 0$.
By choosing the grating strength and detuning, this can be realized at either of the band-edges [see examples in Figs.~\rpict{dispersion}(a) and~(c)], or at a selected band-edge [upper band-edge in Figs.~\rpict{dispersion}(b) and~(d)]. In contrast, if $\eta_j > 0$, the sign of the curvature changes sign and the edges of the band-gap appear at points with non-zero wavenumbers $\pm k_j$, as at the lower gap edges in Figs.~\rpict{dispersion}(b) and~(d).

A key characteristic of slow-light at the band-edges is the second-order dispersion $D = \mathrm{d}^2 \omega / \mathrm{d} k^2$. We find that the absolute value of $D[\omega_j(k=0)]$ is proportional to $|\eta_j|$. Therefore, the dispersion at the gap-edge becomes quartic when $\eta_j=0$ as $\omega_j = \omega_j(k=0) + 0 \times k^2 + O(k^4)$, and this happens precisely at the transition of dispersion curves from the single band-edge with $k_j=0$ to double band-edges with $\pm k_j \ne 0$. The quartic dispersion occurs at the upper gap-edges in Figs.~\rpict{dispersion}(b),(d).

In Fig.~\rpict{tuning}, we present the dependence of the wavenumbers ($k_j$) and second-order dispersion at the band-gap edges. We note that Eqs.~\reqt{BlochEqSymm} and~\reqt{BlochMatrS} possess a general symmetry, $\omega(k; \delta_2) = -\omega(k; -\delta_2)$, allowing us to represent simultaneously the characteristics for both the upper and lower band-edges.
Results of calculations for superimposed Bragg gratings
[Figs.~\rpict{tuning}(a),(b)] or Bragg and long-period gratings
[Figs.~\rpict{tuning}(c),(d)] demonstrate that both the wave-number and dispersion of band-edge
slow light can be engineered by choosing the grating detuning ($\delta_2$) and
strength ($\rho_2$), the values of which can be freely selected by specifying
the required grating period and the refractive index contrast. Consistent with
the theoretical predictions, the absolute value of $D$ is reduced to zero
[visible as white stripes in Figs.~\rpict{tuning}(b) and~(d)] along the path
in the $\delta_2 - \rho_2$ plane where band-edge wavenumbers $k_j$ start to deviate from zero.

We note that in Eq.~\reqt{grating} we assumed that the grating planes are
orthogonal to the fiber axes. It was pointed out by Erdogan and Sipe
\cite{Erdogan:1996-296:JOSA} that by tilting these planes the coupling between
selected modes can be enhanced. However, the detailed design of the grating is
outside the scope of this work.

We stress that the technology for writing gratings in optical fibres is mature
and the fabrication of the gratings described here should not be a problem, in
principle. An additional degree of freedom that can be brought to bear in
possible experiments is that the frequency difference between the resonances
associated with the two gratings can be tuned be stretching the fiber.

In conclusion, we have demonstrated that both the phase velocity and the
strength of second-order dispersion in the slow-light regime can be controlled
in optical fibers with super-structured Bragg gratings or Bragg and long-period
gratings designed for coupling of fundamental and higher-order modes. In
particular, it is possible to realize flat band-edge dispersion of quartic
type, when the second-order dispersion is suppressed. As a final point, though
we considered here an optical fiber geometry, any multi-moded guided-wave
geometry in which gratings can be fabricated is suitable for dispersion control of slow-light.

This work has been supported by the Australian Research Council through the
Centre of Excellence CUDOS research projects.

\end{document}